\documentclass[floats,floatfix,showpacs,preprintnumbers,amssymb,prd,twocolumn,superscriptaddress,nofootinbib,nolongbibliography,reprint]{revtex4-1}

\usepackage{amssymb,amsmath,verbatim,mathtools,needspace,enumitem,etoolbox,graphicx,physics,microtype,afterpage,xspace,tabularx,lmodern,multirow,boldline, makecell, booktabs}
\usepackage{gensymb}
\usepackage{appendix}
\usepackage{tensor}
\usepackage{array}
\usepackage{siunitx}
\usepackage[normalem]{ulem}
\usepackage[dvipsnames, usenames]{xcolor}
\usepackage{xr-hyper}
\definecolor{linkcolor}{rgb}{0.0,0.3,0.5}
\usepackage[unicode, colorlinks=true, linkcolor=linkcolor, citecolor=linkcolor, filecolor=linkcolor, urlcolor=linkcolor, linktocpage, breaklinks]{hyperref}
\usepackage[all]{hypcap}
\usepackage[T1]{fontenc}
\usepackage[utf8]{inputenc}
\usepackage[usenames,dvipsnames]{xcolor}
\hypersetup{colorlinks=true,citecolor=magenta,linkcolor=violet,urlcolor=magenta}
\usepackage{multirow}

\definecolor{romared}{RGB}{142,0,28}

\newcommand{\be}{\begin{equation}}
\newcommand{\ee}{\end{equation}}

\def\be{\begin{equation}}
\def\ee{\end{equation}}
\newcommand{\beq}{\begin{eqnarray}}
\newcommand{\eeq}{\end{eqnarray}}

\usepackage{makecell}
\usepackage{soul}

\newcolumntype{Y}{>{\centering\arraybackslash}X}

\makeatletter
\newcommand*{\addFileDependency}[1]{
  \typeout{(#1)}
  \@addtofilelist{#1}
  \IfFileExists{#1}{}{\typeout{No file #1.}}
}
\makeatother

\newcommand*{\myexternaldocument}[1]{%
    \externaldocument{#1}%
    \addFileDependency{#1.tex}%
    \addFileDependency{#1.aux}%
}

\myexternaldocument{Supp}

\begin{document}

\title{On the non-zero Love numbers of magnetic black holes}

\author{David Pere\~niguez}
\email[]{dpereni1@jhu.edu}
\affiliation{William H. Miller III Department of Physics and Astronomy, Johns Hopkins
University, 3400 North Charles Street, Baltimore, Maryland, 21218, USA}
\affiliation{Center of Gravity, Niels Bohr Institute, Blegdamsvej 17, 2100, Copenhagen, Denmark}
\author{Edgars Karnickis}
\email[]{edgars.karnickis@gmail.com}
\affiliation{Center of Gravity, Niels Bohr Institute, Blegdamsvej 17, 2100, Copenhagen, Denmark}
\date{\today}

\begin{abstract}
Black holes are believed to possess vanishing Love numbers, which implies that they do not deform in the presence of external tides. This fact has been verified in a number of scenarios, that involve tides of bosonic fields of various natures (e.g. scalar, electromagnetic and gravitational), and has triggered active research in trying to identify the underlying reason. Surprisingly, two counter-examples have been found recently. The first, concerns charged-field tides on electrically-charged black holes. In that case, however, the response cannot be disentangled from dissipative effects, and might thus be argued to not consist of a truly conservative deformation. The second concerns fermionic tides on neutral holes. While these yield a purely conservative response, they lack a classical interpretation, which is the physical regime where black hole deformability is understood. Here, we consider magnetic Reissner--Nordström black holes and show that electrically-charged, scalar-field tides induce non-vanishing tidal Love numbers. We prove that this is a purely non-dissipative effect, in contrast to the cases of rotating or electrically-charged black holes, and hence consists of a genuine deformation. In addition, the magnetic charge resolves common ambiguities in defining Love numbers, so our result does not rely on any regularisation scheme. This constitutes a clear realisation of how new physics can influence black hole tidal deformability, and offers new perspectives on the study of black hole Love numbers.
\end{abstract}

\maketitle


\noindent{\bf{ Introduction.}} 
The response of a self-gravitating system to the presence of a companion is encoded in the so-called tidal Love numbers (TLNs) \cite{Love1,Love2,Love3}. They depend only on the properties of the deformed object and, therefore, encode information about its intrinsic structure. While most bodies such as stars are deformed under the action of an external tide, with important consequences for gravitational wave physics \cite{Flanagan:2007ix,Cardoso:2017cfl,LIGOScientific:2018cki,Chatziioannou_2020}, black holes are remarkably rigid. Precisely, their TLNs vanish \cite{Binnington:2009bb,Damour:2009vw,Gurlebeck:2015xpa,LeTiec:2020spy,LeTiec:2020bos,Chia:2020yla}, a fact that has been argued to hold even at non-linear level \cite{De_Luca_2023,Riva:2023rcm,Iteanu:2024dvx,Kehagias:2024rtz,Combaluzier-Szteinsznaider:2024sgb,Gounis:2024hcm},
has been related to symmetry arguments \cite{Bhatt:2023zsy,Charalambous:2021kcz,Hui:2021vcv,Berens:2022ebl,BenAchour:2022uqo,Charalambous:2022rre,Katagiri:2022vyz,Rai:2024lho,Sharma:2024hlz,lupsasca2025loveblackholes}, and it can be studied from several perspectives, such as the effective field theory framework \cite{Kol:2011vg,Porto:2016zng,Hui:2020xxx,Ivanov:2022qqt,Bonelli:2021uvf,Charalambous:2021mea,Ivanov:2022hlo,Glazer:2024eyi}, as well as Hamiltonian and scattering approaches \cite{Steinhoff:2016rfi,Sennett:2017etc,Creci:2021rkz,Gupta:2020lnv}. 

Black holes exhibit non-trivial TLNs in certain scenarios, such as in theories beyond general relativity \cite{Cardoso:2018ptl,DeLuca:2024tlb,DeLuca:2022bjs,Barbosa:2025run,Barura:2024tln}, in higher spacetime dimensions \cite{Kol:2011vg,Hui:2020xxx,Chakravarti:2018vlt,Chakravarti:2019aup,Pereniguez:2021xcj,Dey:2020lhq,Dey:2020kdp,Cardoso:2019wbh,Rodriguez:2023rqp,Charalambous:2023vcy,Charalambous:2024pqe,Charalambous:2024thesis,Singha:2024tld} or in (A)dS spacetimes \cite{Emparan:2017qxd,Nair:2024dS,Franzin:2024ads,Bhatt:2024mvr}. Certain composite black hole-matter systems can also present nontrivial TLNs, like black holes dressed with superradiant boson clouds \cite{Baumann:2018vus,DeLuca:2021ite,DeLuca:2022xlz,Brito:2023pyl} or some effective models of matter \cite{Cardoso:2019upw,Cannizzaro:2024fpz}. 
However, when considering the genuine asymptotically-flat black hole solutions of four-dimensional general relativity, possibly charged under fundamental gauge-fields (e.g.~the electromagnetic field) the vanishing of TLNs seems a robust and fascinating property \cite{Cardoso:2017cfl,Poisson:2021yau,Pereniguez:2021xcj,Cvetic:2021vxa,Charalambous:2024tdj,Rai:2024lho}. 

In two interesting recent works \cite{Ma:2024few,Chakraborty:2025zyb}, it has been observed that this is not always true. In \cite{Ma:2024few} the authors show that charged-scalar static tides induce a non-trivial response on electrically-charged black holes. Although this is surprising, such response cannot be disentangled from dissipative effects (as we show below), and it might be argued that it does not correspond to a genuine deformation. This is reminiscent of the case of rotating black holes \cite{LeTiec:2020spy,Chia:2020yla}, where the response is purely due to dissipation. On the other hand, in \cite{Chakraborty:2025zyb} the authors note that Kerr black holes turn out to possess non-zero \textit{fermionic} TLNs. These appear to be real numbers even if the hole is in a rotating state, which indicates that they are true deformations with no dissipation. Physically, such absence of dissipation can be understood from the fact that fermions do not feel stimulated superradiance, mainly due to their lack of a classical interpretation, by Pauli's principle (see e.g.~\cite{Brito:2015oca}). However, the question of black hole tidal deformability is a classical-physics one, so it might still be argued that black holes do not deform.

In this work we consider a bosonic system, consisting in a minimally-coupled, electrically-charged scalar field on the background of a magnetic Kerr--Newman (KN) black hole, which is asymptotically flat and determined by its mass $M$, angular momentum $J$ and magnetic charge $P$. This situation has been less explored in the literature than the case of an electrically-charged black hole. While experimental evidence suggests magnetic monopoles and magnetic black holes likely do not exist \cite{Giacomelli:2003yu,ATLAS:2023esy,Workman:2022ynf}, they are physically sound at a theoretical level \cite{tHooft:1974kcl,Polyakov:1974ek,Ortin:2015hya} and could have been produced in the early universe or at late stages of Hawking evaporation \cite{Preskill:1984gd,PhysRevD.15.3530,Stojkovic:2004hz,Profumo:2024fxq}. Besides, at a conceptual level the physics of magnetic monopoles and black holes is very rich, and has applications to understand black hole stability \cite{Gibbons:1990um,PhysRevD.45.R2586,Lee:1991vy,Lee:1991qs,Pereniguez:2024fkn} beyond-Standard Model physics \cite{Maldacena:2020skw,Bai:2020spd,Bai:2020ezy,Gervalle:2024yxj}, gravitational waves \cite{Liu:2020vsy,Liu:2020bag,Carullo:2021oxn,Chen:2022qvg,Pereniguez:2023wxf,Dyson:2023ujk,DeFelice:2023rra,Grilli:2024fds}, dark matter and cosmology \cite{Turner:1982ag,Bai:2019zcd,Kritos:2021nsf,Diamond:2021scl,Kobayashi:2023ryr,Zhang:2023zmb,Wang:2023qxj}, and even astrophysical plasmas \cite{Gralla:2014yja}.

In the specific case considered here, that of a charged scalar evolving on a magnetic KN background, it has been shown \cite{Pereniguez:2024fkn} that phenomena such as the superradiant instability exhibits features absent in neutral or electric black holes, with implications for new hairy solutions \cite{Cunha:2024gke}. It is thus natural to explore whether black hole tidal deformability also presents novel properties in this context. When studying the effect of a scalar tide we find that, thanks to the hole's magnetic charge, the static response coefficient $\mathcal{F}_{\ell,m}$ \cite{Chia:2020yla,Chakraborty:2025zyb} can be defined unambiguously. We provide an analytic expression for it, and show that it scales with the black hole temperature $T_{H}$ as
\begin{equation}
    \mathcal{F}_{\ell,m}\sim T_{H}^{\sqrt{(1+2\ell)^{2}-N^{2}}}\, ,\quad \left(\ell\geq\lvert N\rvert/2\right)\,,
\end{equation}
where $\ell$ is the tidal Wu--Yang harmonic number (introduced below) and $N$ is the number of magnetic monopoles inside the black hole. In the rotating case, $\mathcal{F}_{\ell,m}$ presents a real and an imaginary part. The latter is induced by the fluxes of conserved charges across the horizon, which we compute explicitly. In the non-rotating case such fluxes vanish, and so does the imaginary part of the response coefficient. However, its real part remains non-zero, thus yielding a non--vanishing TLN. Unlike fermionic TLNs, it vanishes at extremality when $T_{H}\to0$, similarly to what was observed in \cite{Pereniguez:2021xcj} for higher-dimensional charged black holes. 

This paper is structured as follows. First, we describe some relevant aspects about massive, electrically-charged scalars evolving on dyonic KN black holes. Next, we restrict to massless, charged scalars and a purely magnetic KN background and compute the scalar response to a stationary tidal field, showing that in the non-rotating case the TLNs numbers do not vanish. We conclude with a brief discussion of our results and of potential future directions.   

\noindent{\bf{ Charged fields on dyonic KN black holes.}} 
The action describing a charged scalar field minimally-coupled to the Einstein--Maxwell theory is ($G=c=\hbar=1$)
\begin{align} \notag
S[g,A,\varphi]&=\frac{1}{16\pi}\int d^{4}x\sqrt{-g}\left(R-F^{2}\right)\\ 
&-\frac{1}{2} \int d^{4}x\sqrt{-g}\left(D_{a}\bar{\varphi}D^{a}\varphi+\mu^{2}\bar{\varphi}\varphi\right)\, , \label{Action}
\end{align}
where $D_{a}=\nabla_{a}+ieA_{a}$ is the gauge-covariant derivative and $e$ and $\mu$ are the charge and the mass of the scalar field. The theory is invariant under a $U(1)$-gauge symmetry generated by a local real function $\alpha(x)$,  
\begin{equation}\label{eq:gauge}
\varphi \mapsto e^{i\alpha(x)}\varphi\, , \ \ \ \ \ A_{a}\mapsto A_{a}-\frac{1}{e}\ \nabla_{a}\alpha(x)\, .
\end{equation}
All solutions to the Einstein--Maxwell theory are solutions to \eqref{Action} with a vanishing scalar profile. Fluctuations about those backgrounds include metric $\delta g_{\mu\nu}$, electromagnetic $\delta A_{\mu}$ and scalar perturbations $\delta\varphi$. However, due to the vanishing scalar background, first-order scalar fluctuations decouple from the gravitational and electromagnetic ones, and satisfy (henceforth we write $\delta\varphi\equiv
\varphi$ to simplify notation),
\begin{equation}\label{eq:waveeq}
    \left(D^{\mu}D_{\mu}-\mu^{2}\right)\varphi=0\, .
\end{equation}
The most general asymptotically flat black hole in the Einstein--Maxwell theory is the dyonic KN solution. It describes a hole of mass $M$, angular momentum $J$, and electric and magnetic charges $Q,P$. The fundamental field's charge $e$ and the hole's magnetic charge are constrained by Dirac's quantisation condition \cite{Dirac:1931kp},
\begin{equation}
    2eP=N\,, \quad N=0,\pm1,\pm2,...
\end{equation}
Unlike the quantisation of electric charges, Dirac's quantisation condition must hold at the classical level (the theory is otherwise not well-defined geometrically). This makes the analysis of electrically-charged fields on magnetic backgrounds qualitatively different from the purely electric case. This is described in detail in \cite{Pereniguez:2024fkn}, for electric scalars \eqref{eq:waveeq} on the dyonic KN background where all explicit equations can be found.\footnote{In particular, we remind the Maxwell potential in the \textit{intermediate gauge} reads $ A=-\frac{Q r}{\Sigma}\left(dt-a \sin^{2}\theta d\phi\right)+\frac{P\cos\theta}{\Sigma} \left(a dt-(r^2+a^2)d\phi\right)$, with $\Sigma=r^{2}+a^{2}\cos^{2}\theta$.} Here we follow that reference and simply highlight the aspects that will be relevant later on. Mode solutions to equation \eqref{eq:waveeq}, in the so-called intermediate gauge, have the form
\begin{equation}\label{eq:modes}
    \varphi=e^{-i\omega t}R(r)Y_{N,\ell,m}\left(\theta,\phi\right)=e^{-i\omega v}\mathcal{R}(r)Y_{N,\ell,m}\left(\theta,\chi\right)\,,
\end{equation}
where the first expression is written in terms of Boyer--Lindquist coordinates and the second in advanced Eddington--Finkelstein ones (notice the radial factors differ). The angular dependence is encapsulated in the Wu--Yang monopole harmonics $Y_{N,\ell,m}$ \cite{Wu:1976ge} (also reviewed in \cite{Pereniguez:2024fkn}),\footnote{Strictly speaking, they are a smooth deformation of the original Wu--Yang monopole harmonics, due to the black hole spin $a=J/M$. However, for static, massless perturbations $\omega=\mu=0$, they are precisely the Wu--Yang monopole harmonics.} whose quantum angular numbers $\ell,m$ satisfy
\begin{equation}\label{eq:quantumang}
    \ell=\frac{\vert N\rvert}{2},\frac{\vert N\rvert}{2}+1...\, ,\quad  m=-\ell,-\ell+1,...,\ell-1,\ell\, .
\end{equation}
We note that, in general, these can take both integer and half-integer values. Besides, monopole harmonics present structural differences relative to the standard harmonics (see \cite{Pereniguez:2024fkn}). The symmetries of the dyonic KN background entail the existence of conserved gauge-invariant currents describing the energy and angular momentum carried by the scalar, and associated to the $U(1)$-gauge symmetry is the usual conserved electric-charge current. For a mode of the form \eqref{eq:modes}, the fluxes of energy, angular momentum and charge across the event horizon $H$ at $r=r_{+}$ are \cite{Pereniguez:2024fkn}
\begin{equation}\label{eq:fluxes}
    \begin{aligned}
    \Delta E=&\left(\omega_{I}^{2}+\omega_{R}\omega_{*}\right)\lvert \mathcal{R}(r_{+})\rvert^{2}\int_{H}e^{2 \omega_{I} v}\lvert  Y_{N,\ell,m}\left(\theta,\chi\right)\rvert^{2} \tilde{\boldsymbol{\epsilon}}\, ,\\ 
    \Delta J=&m\omega_{*}\lvert \mathcal{R}(r_{+})\rvert^{2}\int_{H}e^{2 \omega_{I} v}\lvert  Y_{N,\ell,m}\left(\theta,\chi\right)\rvert^{2}\tilde{\boldsymbol{\epsilon}}\,\, ,\\ 
    \Delta Q=&-e \omega_{*}\lvert \mathcal{R}(r_{+})\rvert^{2}\int_{H}e^{2 \omega_{I} v}\lvert Y_{N,\ell,m}\left(\theta,\chi\right)\rvert^{2}\tilde{\boldsymbol{\epsilon}}\, ,
\end{aligned}
\end{equation}
where we separate the real and imaginary parts of the frequency $\omega=\omega_{R}+i\omega_{I}$, and $\tilde{\boldsymbol{\epsilon}}$ denotes the natural volume form on $H$ relative to the future-directed generator of the horizon. The superradiant threshold frequency $\omega_{*}$ is given by
\begin{equation}
    \omega_{*}=\omega_{R}-m\Omega+e\Phi\,,
\end{equation}
with $\Omega=a/(r_{+}^{2}+a^{2})$ and $\Phi=r_{+}Q/(r_{+}^{2}+a^{2})$ the black hole's angular velocity and electric potential, respectively, and $a=J/M$. The integrals in \eqref{eq:fluxes} are positive-definite, so the prefactors indicate the sign of the flux. These formulas, valid for the most general scalar mode \eqref{eq:modes}, will be useful in interpreting the static solutions constructed in the next section.

\noindent{\bf{ Static tidal fields and response coefficients.}} So far the discussion has been general, considering a dyonic KN background and a scalar with non-vanishing mass $\mu$ and charge $e$. Here, we shall restrict to purely magnetic KN black holes $Q=0$, with $P\ne0$ and $a\ne0$, and charged, massless fields $e=N/2P\ne0$ and $\mu=0$. Our goal is to obtain the response of the scalar to an external, stationary tidal field. This is captured in the static response coefficient $\mathcal{F}_{\ell,m}$, defined as follows \cite{Binnington:2009bb,Chia:2020yla}. An asymptotic analysis of the equation satisfied by $R(r)$ in \eqref{eq:modes} shows that, for $r/r_{+}\gg1$, the most general stationary solution ($\omega=0$)\footnote{Notice that $\pounds_{\partial_{t}}\varphi=0$ is not a good notion of stationarity since it is not gauge-invariant. A charged scalar $\varphi$ is stationary relative to a timelike Killing vector $K$ if $\delta_{K}\varphi\equiv-K^{a}D_{a}\varphi+i e \mathcal{P}_{K}\varphi=0$, where $\mathcal{P}_{K}$ is a function satisfying $\nabla_{a}\mathcal{P}_{K}+K^{b}F_{ba}=0$, known as momentum-map \cite{Elgood:2020svt,Ortin:2022uxa,Ballesteros:2023iqb}. It turns out that the quantity $\delta_{\partial_{t}}\varphi$, when written in our gauge for a mode \eqref{eq:modes} and employing the dyonic KN momentum maps \cite{Pereniguez:2024fkn}, gives $\delta_{\partial_{t}}\varphi=-\pounds_{\partial_{t}}\varphi=i\omega\varphi$ so $\delta_{\partial_{t}}\varphi=0$ implies $\omega=0$.} has the form
\begin{equation}\label{eq:tidalsol}
    R(r)=\mathcal{E}\left(\frac{r}{r_{+}}\right)^{\frac{L-1}{2}}\left[R_{\text{tide}}(r)+\mathcal{F}_{\ell,m} \left(\frac{r_{+}}{r}\right)^{L}R_{\text{resp}}(r)\right]\, .
\end{equation}
Here, we introduced the quantity
\begin{equation}
    L\equiv\sqrt{(1+2\ell)^{2}-N^{2}}\, ,
\end{equation}
the radial functions are normalised as $R_{\text{tide,resp}}(r)=1+O(1/r)$, $\mathcal{E}$ is an arbitrary constant that measures the strength of the tidal field, and $\mathcal{F}_{\ell,m}$ is at this point an arbitrary constant. Each term in the brackets of \eqref{eq:tidalsol} corresponds to a solution and, if the number of monopoles is different from zero $N\ne0$, they are manifestly linearly independent since $L$ is in general not an integer (we will comment on exceptions below). This fact allows us to unambiguously write the general solution as a tidal contribution $\sim \left(r/r_{+}\right)^{\frac{L-1}{2}}R_{\text{tide}}(r)$, which grows with $r$ and is generated by an external source, and a decaying solution $\sim \left(r_{+}/r\right)^{\frac{1+L}{2}}R_{\text{resp}}(r)$ which is interpreted as the stationary response of the system. The magnitude of the response is measured by the coefficient $\mathcal{F}_{\ell,m}$, which has been unspecified so far. In our case, it is fixed by requiring that the solution is regular at the event horizon, which determines it uniquely in terms of the black hole parameters. In particular, it is independent of the external tide's amplitude $\mathcal{E}$ and so it encodes information about the intrinsic structure of the system. To obtain $\mathcal{F}_{\ell,m}$ in this way, we notice that upon the redefinitions
\begin{equation}
    R(r)=\left(\frac{r-r_{+}}{r-r_{-}}\right)^{i\frac{m\Omega}{2\kappa_{+}}}\left(\frac{r_{+}-r_{-}}{r-r_{-}}\right)^{\frac{1+L}{2}}F\left(\frac{r-r_{+}}{r-r_{-}}\right)\,,
\end{equation}
where $\kappa_{+}=(r_{+}-r_{-})/(2(r_{+}^{2}+a^{2}))$ is the horizon's surface gravity, then the function $F(z)$ with $z \equiv (r-r_{+})/(r-r_{-})$ satisfies the hypergeometric equation with coefficients
\begin{equation}
    a=\frac{1+L}{2}+i m \frac{\Omega}{\kappa_{+}}\, , \ \ b=\frac{1+L}{2}\, ,\ \  c=1+a-b\, .
\end{equation}
Regularity at the event horizon requires that $F(z)$ is itself regular at $z=0$ (see \cite{Pereniguez:2024fkn}), so the solution is fixed to 
\begin{equation}\label{eq:Fsol}
F(z)=\mathcal{A} \ \bold{F}\left(a,b,c;z\right)\, ,
\end{equation}
where $\bold{F}\left(a,b,c;z\right)$ is the hypergeometric function and $\mathcal{A}$ is an arbitrary constant. The other independent solution close to the horizon behaves as $\sim z^{1-c}=z^{-i m \Omega/\kappa_{+}}$, which is singular if $\Omega\ne0$. When $\Omega=0$, such second solution contains a logarithm $\sim\ln z$ as follows form ODE theory and should be discarded too for regularity reasons, so in all cases one is left with \eqref{eq:Fsol}. The domain of definition extends from $z=0$ (horizon) to $z=1$ (infinity) so using the appropriate connection formula,
\begin{widetext}
\begin{equation}
\begin{aligned}
    \bold{F}\left(a,b,c;z\right)&=\frac{\Gamma(c)\Gamma(c-a-b)}{\Gamma(c-a)\Gamma(c-b)}\bold{F}\left(a,b,1+a+b-c;1-z\right)\\
    &+\frac{\Gamma(c)\Gamma(a+b-c)}{\Gamma(a)\Gamma(b)}(1-z)^{c-a-b}\bold{F}\left(c-a,c-b,1+c-a-b;1-z\right)\, ,
\end{aligned}    
\end{equation}
the solution is easily cast in the form \eqref{eq:tidalsol}, which yields
\begin{equation}\label{eq:resp}
    \begin{aligned}
\mathcal{F}_{\ell,m}=&\pi^{-2}\left(\frac{r_{+}-r_{-}}{r_{+}}\right)^{L}\bigl\lvert \Gamma \left[(1+L)/2\right]\bigr\rvert^{2}\bigl\lvert \Gamma \left[(1+L)/2+im\Omega/\kappa_{+}\right]\bigr\rvert^{2}\left[\Gamma\left(-L\right)/\Gamma\left(L\right)\right]\cos\left(\pi L/2\right)\\
&\times\left[\cosh{\left(\pi m \Omega/\kappa_{+}\right)}\cos{\left(\pi L/2\right)}+i \sinh{\left(\pi m \Omega/\kappa_{+}\right)}\sin{\left(\pi L/2\right)}\right]\\
\equiv& \kappa_{N\ell m}+i \nu_{\ell m}\, .
\end{aligned}
\end{equation}
\end{widetext}
First, it is worth to emphasize that this formula is well defined whenever the parameters take their physical values, that is, $N\in\mathbb{Z}$ and $\ell,m$ given by \eqref{eq:quantumang}. This is in contrast with the neutral case $N=0$, where $L\to1+2\ell$ becomes an odd integer and the factor $\Gamma\left(-L\right)\cos\left(\pi L/2\right)$ is ill-defined. A common practice in the literature consists in performing an ``analytic continuation in $\ell$'', meaning promoting $\ell$ to be a real number and subsequently taking the limit $\ell\to\mathbb{N}$. Our formula coincides, in the neutral case $N=0$, with the $\ell$-analytically-continued response coefficient obtained previously in the literature (see e.g. \cite{Chakraborty:2025zyb}). In the charged case $N\ne0$, however, this kind of inconvenience is absent and the response coefficient \eqref{eq:resp} yields a sensible answer directly. An exception are the cases where $L$ is natural, that is, when the numbers $1+2\ell$ and $N$ are the hypotenuse and a leg of a Pythagorean triple. In that case, the asymptotic form of the solution is not \eqref{eq:tidalsol} any more, and it either exhibits a logarithmic tail yielding a ``running'' coupling, or the solution is an algebraic function and TLNs vanish. We notice this mimics the situation encountered for higher-dimensional black holes, which we will discuss below. A second aspect to highlight is that \eqref{eq:resp} exhibits a real ($\kappa_{N\ell m}$) and an imaginary part ($\nu_{\ell m}$) (we notice the first line in \eqref{eq:resp} is real due to the property $\bar{\Gamma}(z)=\Gamma(\bar{z})$). The latter is usually associated with dissipation effects \cite{LeTiec:2020spy,Chia:2020yla,LeTiec:2020bos}. This is justified by comparing it with the fluxes across the event horizon, \eqref{eq:fluxes}. For a zero-frequency mode, we find $\Delta E=0$, but $\Delta J\sim\Delta Q\sim m \Omega$. The imaginary part in \eqref{eq:resp} is indeed $\sim \sinh{\pi m \Omega/\kappa_{+}}$, so it vanishes precisely when there is no dissipation, $m \Omega\to0$. However, the most important observation of this work is that, in the case of a non-rotating magnetic black hole $\Omega=0$, where there is no dissipation ($\Delta E=\Delta J=\Delta Q=0$), the real part of the response coefficient is still non-vanishing, and given by
\begin{widetext}
\begin{equation}\label{eq:TLN}
    \kappa_{N\ell m}=\frac{\left(2 r_{+}\kappa_{+}\right)^{\sqrt{(1+2\ell)^{2}-N^{2}}}}{\pi^{2}}
    \cos^{2}\left[(\pi/2) \sqrt{(1+2\ell)^{2}-N^{2}}\right]
    \Bigl\lvert \Gamma \left[\left(1+\sqrt{(1+2\ell)^{2}-N^{2}}\right)/2\right]\Bigr\rvert^{4}\frac{\Gamma\left(-\sqrt{(1+2\ell)^{2}-N^{2}}\right)}{\Gamma\left(\sqrt{(1+2\ell)^{2}-N^{2}}\right)}\, .
\end{equation}    
\end{widetext}
To the best of our knowledge, this is the first example where an isolated, asymptotically-flat black hole within general relativity presents a genuine non-dissipative response to a bosonic tide, encapsulated in a non-vanishing TLN. As mentioned earlier, it has been recently pointed out that \textit{fermionic} black hole TLNs can also be non-vanishing \cite{Chakraborty:2025zyb}. However, the lack of a \textit{classical} interpretation of fermionic fields questions whether this effect might have a relevant impact in the classical dynamics. In the present case, the non-trivial conservative response concerns a boson so, in principle, this effect could play a classical role (although, of course, neither magnetic black holes or massless charged fields are expected to exist, or at least not abound, in nature).
It is also interesting to compare our result to the case of a purely electric ($P=0$) KN black hole, recently considered in \cite{Ma:2024few}. In that case, both real and imaginary parts of the response coefficient are non-vanishing, even in the non-rotating limit $a=0$. This is due to the fact that such response cannot be disentangled form dissipation, since there is always flux through the horizon with $\Delta J\sim\Delta Q\sim e Q$, so it does not consist of a pure deformation. In addition, the limit $e\to0$ is not well defined, potentially due to the breaking of the local $U(1)$ gauge symmetry when $e=0$. As a last observation, unlike the fermionic case our bosonic TLNs vanish at extremality, $\kappa_{\ell m}\to0$ as $T_{H}=\kappa_{+}/2\pi\to0$. The vanishing of scalar TLNs in the extremal regime has been previously observed in higher dimensions \cite{Pereniguez:2021xcj}, where in general black holes present non-trivial tidal deformability. It is an intriguing fact that this is also true in the present four-dimensional example. 


\noindent{\bf{ Discussion.}} We have presented the first example where an asymptotically-flat black hole within general relativity (a magnetic Reissner--Nordström black hole) exhibits non-vanishing bosonic TLNs, and have shown that these correspond to a genuine, non-dissipative deformation. In addition, we have found that such conservative response vanishes in the extremal limit. This result is free from the ambiguities that usually concern TLNs, and in particular it relies on no regularisation procedures such as analytic continuation in the harmonic numbers. While the system is conceptually well-defined, it is not expected to be relevant in astrophysics, since it involves a magnetic black hole and a charged, massless field. However, it provides an original realisation of how black hole tidal deformability can yield signatures of new physics. In particular, the black holes of some fundamental theories (such as string theory and its low-energy supergravity effective actions \cite{Cvetic:2021vxa}, or simply black holes coupled to Standard Model fields \cite{Bai:2020ezy,Gervalle:2024yxj}) carry magnetic charges, and this work paves the way towards exploring their potentially non-trivial deformability properties. Interestingly, there exist well-defined horizonless compact objects such as Topological Stars \cite{Bah:2020ogh}, which are higher-dimensional and are supported thanks to a magnetic charge. A straightforward computation not shown here reveals that, surprisingly, their TLNs $\kappa_{\ell m}(TS)$ are \textit{identical} to those of a magnetic RN black hole $\kappa_{\ell m}(BH)$ given by \eqref{eq:TLN}, up to an overall factor related to the freedom in choosing units, that is, $\kappa_{\ell m}(TS)\sim\kappa_{\ell m}(BH)$ (this generalises previous results found in \cite{Bianchi:2023sfs} for neutral fields on Topological Stars). This degeneracy shows that some black holes cannot be told apart from horizonless compact objects by their stationary deformability properties. It would be interesting to explore further this degeneracy in the case of other charged, regular compact objects \cite{BeltranJimenez:2022hvs}. 

Our TLNs present significant similarities with higher-dimensional set ups \cite{Kol:2011vg}: (i) they vanish as  $T_{H}\to0$ \cite{Pereniguez:2021xcj} (ii) they are non-zero for generic (but regularity-compatible) values of $\ell$, but in some cases (when $L\in\mathbb{N}$), the solutions either exhibit logarithmic tails or have vanishing TLNs. This fact suggests our results may be describable via a dimensional reduction of a neutral solution. This also sheds light on how our TLNs should avoid the symmetry arguments in the literature \cite{Hui:2021vcv}: generic modes do not belong to a highest-weight representation of $SL(2,\mathbb{R})$, so their TLNs are not zero. However, the symmetry argument might apply to some of the modes, when $1+2\ell$ and $N$ are the hypotenuse and a leg of a Pythagorean triple. We notice this is also the usual structure found in other scenarios, such as analogue black holes \cite{DeLuca:2024nih}. It would also be interesting to understand whether the non-vanishing of fermionic TLNs  \cite{Chakraborty:2025zyb} is related to our bosonic result \eqref{eq:TLN}, presumably by embedding the solution in a supersymmetric set up. Finally, it would be useful to match our results to a world-line EFT description, as this would clarify the physical interpretation. For electrically charged BHs this matching was analyzed in detail in \cite{Rai:2024lho}, where the leading term in the expansion is the world-line action of a point charge. In the present case the situation is more involved: a point magnetic charge does not couple to the Maxwell field through that action. The corresponding EFT therefore appears to require a qualitatively different setup, which we leave for future work.


\noindent{\bf{ Acknowledgements.}} DP thanks Pierre Heidmann and Paolo Pani for insightfull conversations and useful correspondence, and Bogdan Ganchev and Vitor Cardoso for stimulating discussions. We acknowledge financial support by the VILLUM Foundation (grant
no. VIL37766) and the DNRF Chair program (grant
no. DNRF162) by the Danish National Research Foundation. The Center of Gravity is a Center of Excellence funded by the Danish National Research Foundation under grant No. 184. This project has received funding from the European
Union’s Horizon 2020 research and innovation programme
under the Marie Sklodowska-Curie grant agreement No
101007855 and No 101007855. DP is supported by NSF Grants No. AST-2307146, PHY-2513337, PHY-090003, and PHY-20043, by NASA Grant No. 21-ATP21-0010, by John Templeton Foundation Grant No. 62840, by the Simons Foundation, and by Italian Ministry of Foreign Affairs and International Cooperation Grant No. PGR01167.

\bibliography{refMAIN}

\end{document}